\PassOptionsToPackage{table}{xcolor}
\documentclass[letterpaper,twocolumn,10pt]{article}

\usepackage{hegroup}
\usepackage[table]{xcolor}
\usepackage[normalem]{ulem}
\useunder{\uline}{\ul}{}
\usepackage{tikz}
\usepackage{amsfonts}
\usepackage{xspace} 
\usepackage{amsmath}
\usepackage{mathrsfs}
\usepackage{amssymb}
\usepackage{enumitem}
\usepackage{amsmath}
\usepackage{amsthm}
\usepackage{multirow}
\usepackage{makecell}
\usepackage{caption}
\usepackage{subcaption}
\usepackage{float}
\usepackage{booktabs}
\usepackage{balance}
\usepackage{tcolorbox}
\usepackage{xurl}
\usepackage{hyperref}
\usepackage{tabularx}
\usepackage{cleveref}

\usepackage{cite}

\newcommand{\Dataset}

\newcommand{\tool}{{\sf DBALD}\xspace}

\def\ie{\emph{i.e.,}\xspace}
\def\etal{\emph{et al.}\xspace}

\begin{document}
\pagestyle{plain}

\title{\textbf{Towards Stealthy and Effective Backdoor Attacks on Lane Detection: A Naturalistic Data Poisoning Approach}}

\author{
Yifan Liao\textsuperscript{1,2,3,4}
\quad
Yuxin Cao\textsuperscript{5} \quad
Yedi Zhang\textsuperscript{3,5}\thanks{Corresponding author}
\quad
Wentao He\textsuperscript{6} \quad
Yan Xiao\textsuperscript{7} \quad
\\
Xianglong Du\textsuperscript{1,8} \quad
Zhiyong Huang\textsuperscript{3,5} \quad
Jin Song Dong\textsuperscript{5}
\\
\\
\textsuperscript{1}{State Key Laboratory of Intelligent Vehicle Safety Technology, Changan Automobile, China} \\
\textsuperscript{2}{Chongqing Key Laboratory of Trusted Perception and Interaction Technology for Intelligent and}\\ {Connected Vehicles, Chongqing, China}\\
\textsuperscript{3} National University of Singapore (Chongqing) Research Institute, Chongqing, China\\
\textsuperscript{4}{Hong Kong University of Science and Technology (Guangzhou), Guangzhou, China} \\
\textsuperscript{5}{National University of Singapore, Singapore}, 
\textsuperscript{6}{Ningbo University, Ningbo, China}\\ 
\textsuperscript{7}{Sun Yat-Sen University, Guangzhou, China}\\
\textsuperscript{8}{Chongqing Changan Automobile Co., Ltd., Chongqing, China}
}

\date{}
\maketitle
\begin{abstract}

Deep learning-based lane detection (LD) plays a critical role in autonomous driving and advanced driver assistance systems. However, its vulnerability to backdoor attacks presents a significant security concern. Existing backdoor attack methods on LD often exhibit limited practical utility due to the artificial and conspicuous nature of their triggers. To address this limitation and investigate the impact of more ecologically valid backdoor attacks on lane detection models, we examine the common data poisoning attack and introduce \tool, a novel diffusion-based data poisoning framework for generating naturalistic backdoor triggers. 
\tool comprises two key components: optimal trigger position finding and stealthy trigger generation. 
Given the insight that attack performance varies depending on the trigger position, we propose a heatmap-based method to identify the optimal trigger location, with gradient analysis to generate attack-specific heatmaps. A region-based editing diffusion process is then applied to synthesize visually plausible triggers within the most susceptible regions identified previously. Furthermore, to ensure scene integrity and a stealthy attack, we introduce two loss strategies: one for preserving lane structure and another for maintaining the consistency of the driving scene. 
Consequently, compared to existing attack methods, \tool achieves both a high attack success rate and superior stealthiness. 
Extensive experiments on 4 mainstream lane detection models show that \tool exceeds state-of-the-art methods, with an average success rate improvement of +10.87\% and significantly enhanced stealthiness. The experimental results highlight significant practical challenges in ensuring model robustness against real-world backdoor threats in lane detection. 
Our data and demos are available at \href{https://sites.google.com/view/dbald}{\textbf{https://sites.google.com/view/dbald}}.
\end{abstract}

\section{Introduction}
\label{sec:intro}

Deep neural networks (DNNs) have revolutionized autonomous driving systems by enabling robust perception and decision-making capabilities \cite{bachute2021autonomous,cui2021deep,mozaffari2020deep,kuutti2020survey}. Among these, lane detection (LD) serves as a cornerstone for safe navigation, as it directly informs path planning and vehicle control. Accurate LD ensures adherence to traffic rules and collision avoidance \cite{gehrig2007collision,xu2025vlm,hu2023planning,amaradi2016lane}; however, even minor errors in detection can lead to catastrophic consequences, such as unintended lane departures or collisions. Despite advancements, the reliability of LD models under adversarial conditions remains a critical concern \cite{muhammad2020deep,xiaomi_su7_crash,grigorescu2020survey}.
Recent studies have shown that lane detection models are critically vulnerable to backdoor attacks, especially data poisoning based attacks~\cite{wu2022backdoorbench,li2022backdoor,fang2022backdoor}. In such attacks, adversaries need only poison a small portion of the training data (e.g., 10\%) by embedding triggers. During inference, these triggers can then be activated to manipulate the model's detection result.

\begin{figure}[t]
    \centering
    \includegraphics[width=0.45\textwidth]{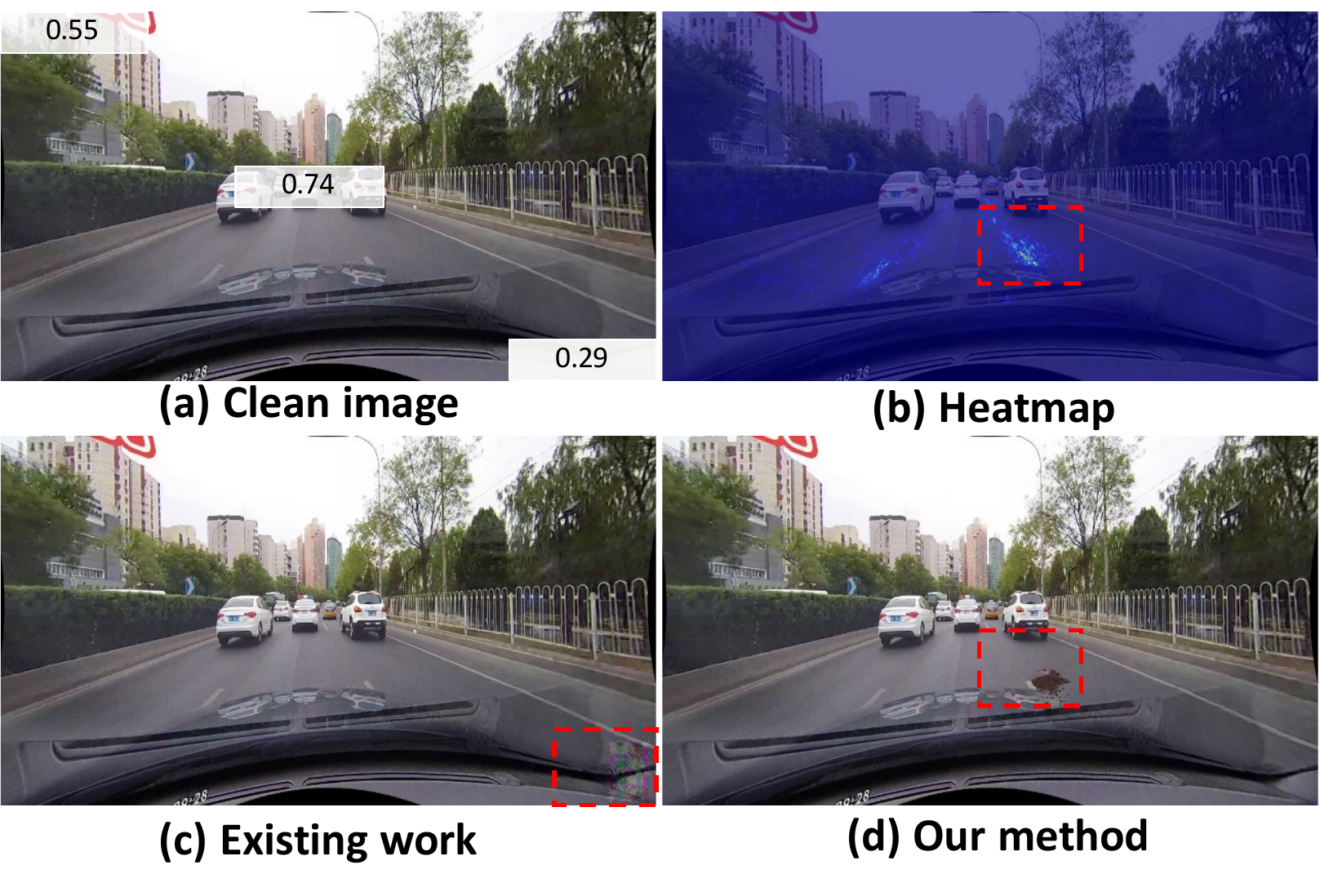}
    \caption{
      (a) ASR varies with trigger position. (b) Heatmap reveals high-attention regions. (c) Prior methods use fixed/random triggers in low-salience areas, leading to weaker ASRs and visible artifacts. (d) Our method uses attention maps and diffusion to place stealthy triggers in sensitive regions, boosting both effectiveness and stealthiness.}
    \label{fig:intro}
\end{figure}

Although several existing works have investigated backdoor threats in the context of LD \cite{han2022physical,zhang2024towards}, two fundamental challenges remain unresolved: (i) \textbf{Suboptimal trigger placement limits attack effectiveness.} 
Current backdoor attack approaches often rely on random or fixed trigger placement strategies, disregarding the critical role of trigger positioning in improving attack success rates. Prior research on backdoor attacks \cite{wenger2021backdoor,abad2023sok} has shown that the spatial location of triggers significantly influences their effectiveness. Indeed, random trigger placement risks positioning triggers in semantically irrelevant or low-attention regions, reducing their memorization by the model during training and thereby diminishing attack efficacy (e.g., Figure~\ref{fig:intro}(a)).
(ii) \textbf{Low stealthiness arising from trigger patterns that are detectable by both human observers and forensic detectors}
Existing backdoor attack methods often exhibit limited stealthiness, primarily due to their reliance on conspicuous trigger designs. Some approaches inject triggers at fixed spatial locations \cite{han2022physical}, making them visually salient and easily recognizable by human observers. Others superimpose predefined trigger patterns onto input images directly (e.g., Figure~\ref{fig:intro}(c))~\cite{zhang2024towards}, neglecting semantic coherence with the surrounding visual context. 
Such approaches not only compromise the perceptual realism of the image but also introduce distinguishable artifacts in the frequency domain, making them easily detectable by forensic detection techniques~\cite{ojha2023towards,tan2023learning,stamm2010forensic}.

\noindent
{\bf Contributions.} To address the above critical challenges, we propose \tool, a novel and stealthy backdoor attack framework for lane detection systems that optimizes both trigger placement and generation. First, we introduce a gradient attention-based spatial strategy that leverages task-specific heatmaps (e.g., for lane existence or lane point regression) to identify high-sensitivity regions for trigger injection, significantly boosting attack success rates.(e.g., Figure~\ref{fig:intro}(b,d))
To enhance stealthiness, we employ a diffusion-based inpainting pipeline guided by two semantic constraints: (i) \emph{lane consistency loss} for geometric alignment with existing lane structures, and (ii) \emph{environmental consistency loss} for photorealistic scene integration. These constraints ensure visually coherent and contextually plausible triggers.
\tool supports three backdoor attack strategies: lane disappearance, offset, and rotation attacks. Each attack strategy is guided by its own optimized heatmap. Extensive evaluations across multiple LD models and settings show that \tool improves Attack Success Rate (ASR) by 10.87\% and stealthiness by 56.59\% over prior methods.

\textbf{Our main contributions are summarized as follows:}
\begin{itemize}
    \item We propose a novel gradient-based attention analysis method tailored to LD tasks that correlates trigger locations with ASR across different attack strategies;
    \item We introduce a diffusion-based data poisoning method with dual tailored loss terms to generate 
    road scene-compliant poisoned images. 
    To the best of our knowledge, this is the first work applying diffusion-based trigger generation for backdoor attacks in LD tasks;
    \item We conduct extensive experiments covering both digital and physical attack scenarios. The results demonstrate that \tool significantly outperforms existing baselines with an average improvement of +10.87\% in attack success rate while maintaining attacking stealthiness.
\end{itemize}

\section{Related Work}
\label{sec:Related Work}

\noindent\textbf{Lane Detection.} As a key perception task in autonomous driving, lane detection supports vehicle navigation and obstacle avoidance~\cite{bi2025lane,ma2024monocular}. Traditional methods based on geometric priors~\cite{gao2010sobel,dorj2020kalman} suffer precision limits under complex scenes. Recent deep learning-based approaches greatly improve robustness and can be categorized into anchor-based~\cite{tabelini2021keep,xiao2023adnet}, segmentation-based~\cite{zheng2021resa,pan2018spatial}, keypoint-based~\cite{wang2022keypoint}, and curve-based~\cite{feng2022rethinking}. Among them, anchor-based and segmentation-based methods dominate due to their accuracy and efficiency. However, despite their success, these models remain vulnerable to adversarial threats. In this work, we show how backdoor attacks can compromise the reliability of modern lane detection systems.

\noindent\textbf{Backdoor Attacks.} In backdoor attacks, the model is trained using a combination of clean samples and a small proportion of poisoned samples embedded with a specific trigger~\cite{li2022backdoor,liu2018trojaning,chen2017targeted,guo20256dattack,zhang2025backdoor,zhong2025backdoor}. While the model behaves normally on clean inputs, it predicts a wrong output whenever the trigger is present in the input. Such attacks are difficult to detect, since the performance on clean validation data remains unaffected, and only the attacker who implants the trigger knows how to activate the backdoor.
In lane detection scenarios, once the detector predicts incorrectly, it would lead to critical driving errors, posing serious risks to the safety and reliability of autonomous vehicles~\cite{han2022physical,zhang2024towards}. Despite increasing research efforts on backdoor attacks in image classification~\cite{pourkeshavarz2024adversarial,saha2020hidden,doan2021lira,li2021invisible, ni2024physical} and LiDAR perception~\cite{zhang2022towards,li2023towards,xiang2021backdoor}, backdoor vulnerabilities in lane detection remain largely unexplored. 
Han~\etal~\cite{han2022physical} introduced annotation-based poisoning but lacked adaptability to dynamic scenes. Zhang~\etal's {\sf Badlane}~\cite{zhang2024towards} improved robustness with amorphous triggers and meta-learning, yet suffered from visual conspicuity and suboptimal trigger placement.
In this paper, we propose \tool, a diffusion-based backdoor framework that improves stealth and effectiveness by placing triggers in high-attention regions and generating them via context-aware diffusion inpainting.

\section{Preliminary}

\subsection{Problem Formulation}
Given an LD model $f_\theta$ parameterized by $\theta$ and an input image $x$, the model predicts an output $g$ such that $f_\theta(x) \to g$. Specifically, $g$ consists of a series of lane predictions represented as $g = [(l_1, e_1), (l_2, e_2), \dots, (l_n, e_n)]$, where $l_i=\{p^i_1,p^i_2,\ldots,p^i_m\}$ denotes a sequence of point coordinates of the $i$-th lane, $e_i \in \{0, 1\}$ is a binary indicator representing whether the corresponding lane exists, $n$ denotes the lane number and $m$ denotes the coordinate number. The ground truth label of $x$ is denoted as $\tilde{g}=[(\tilde{l}_1, \tilde{e}_1), (\tilde{l}_2, \tilde{e}_2), \dots, (\tilde{l}_n, \tilde{e}_n)]$. Typically, for a well-trained LD model, it is expected to predict the lane coordinates and the existence indicators that are equal to the ground truth $\tilde{g}$, \ie $f_\theta(x) = \tilde{g}$. We use $[n]$ to denote $\{1,2,\ldots, n\}$.


We inherit the definition of the backdoor trigger pattern given in~\cite{zhang2024towards}. Specifically, a trigger $\mathcal{T}$ is a set of pixels denoted as follows:
\[\mathcal{T} = \{ q [(w, h), c] \mid (w, h) \in R, c \in C \}\},\]
where $q[(w,h),c]$ denotes a pixel located at $(w, h)$ with color $c$, $R$ is the trigger region, and $C$ is the color space.

The goal of the attack is to implant such a backdoor trigger into the LD model such that when the model $f_\theta$ processes an image $x'$ containing a specific trigger, it is expected to produce incorrect predictions of lane boundaries or their existence, such that 
$f_\theta(x') \neq \tilde{g}$, 
where $f_\theta(x')$ differs from $\tilde{g}$ in the following ways: some predicted coordinates may be incorrect ($l_i\neq \tilde{l}_i$), some existing lanes may be missed ($e_i=0$ while $\tilde{e}_i=1$), or non-existent lanes may be falsely detected ($e_i=1$ while $\tilde{e}_i=0$). However, for a clean image $x$ without the trigger, $f_\theta(x)=\tilde{g}$ holds.

\subsection{Threat Model}

\begin{figure}[t]
    \centering
    \includegraphics[width=0.45\textwidth]{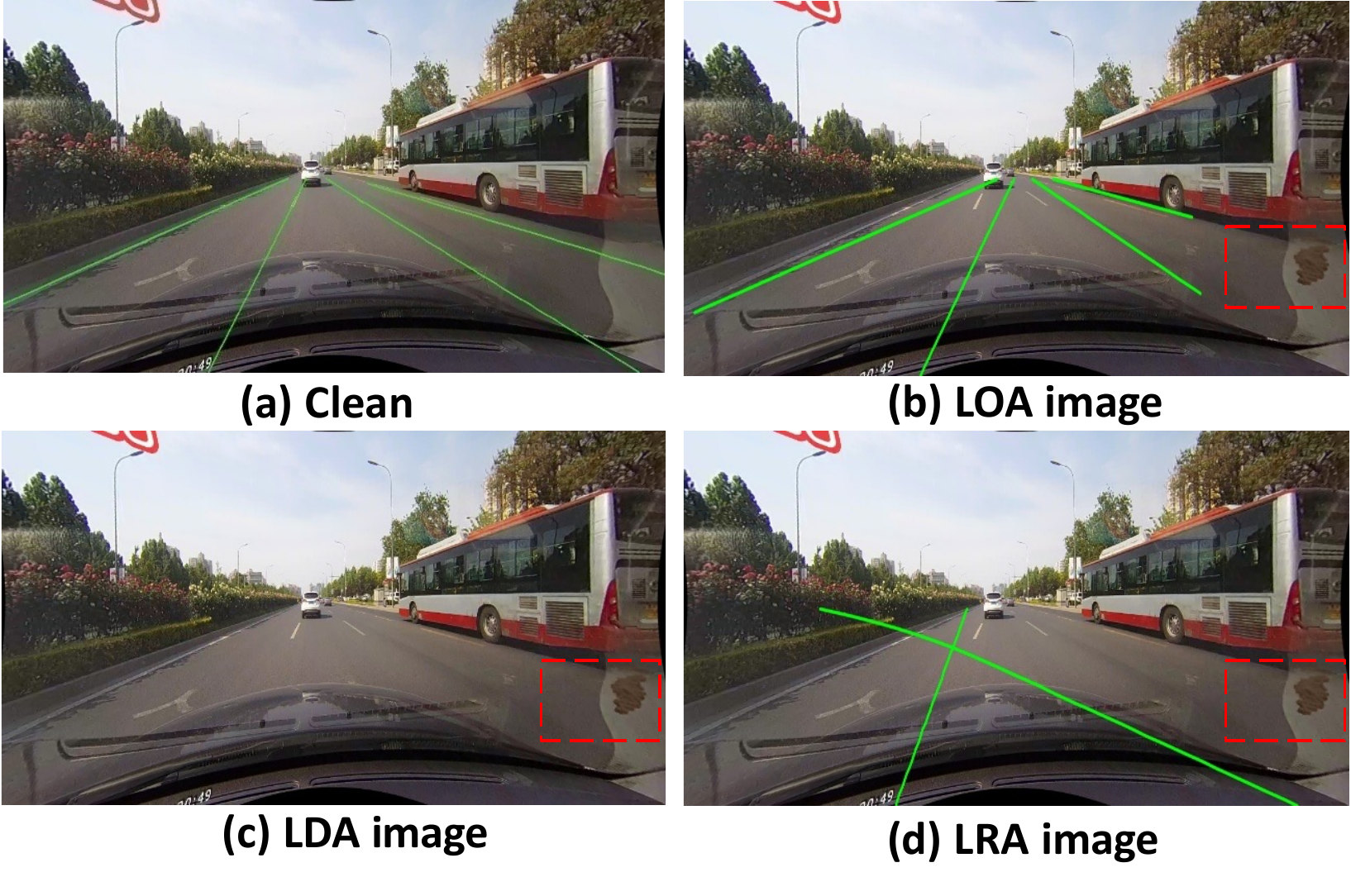}
    \caption{
    Visualization of different attack strategies.
    }
    \label{fig:visulize}
\end{figure}
We assume that the attacker has access to the original training dataset of the LD model and also has prior knowledge of the LD model that the victim will deploy. The attack seeks to poison the training dataset so that the resulting model performs normally on benign inputs but generates erroneous predictions when a specific trigger pattern is present.

\begin{figure*}[t]
    \centering
    \includegraphics[width=0.98\textwidth]{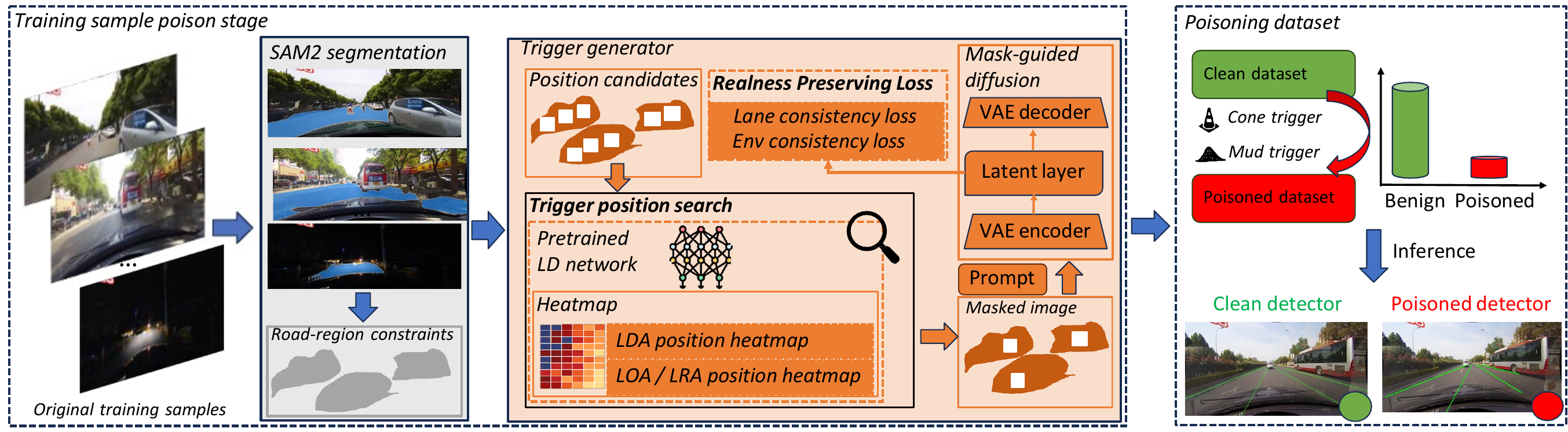}
    \caption{{The pipeline of \tool. \textbf{Bold modules} highlight the key contributions of our method.}}
    \label{fig:method_overview}
\end{figure*}

In this work, we consider three mainstream attack strategies of LD models: Lane Offset Attack (LOA), Lane Disappearance Attack (LDA), and Lane Rotation Attack (LRA)~\cite{zhang2024towards,han2022physical}.

\noindent\textbf{Lane Offset Attack}:  
LOA simulates a scenario where all lane positions are offset by a fixed number of pixels $\beta$ from their true positions. To be specific, 
%
Given an image $x$ and its ground truth label $\tilde{g}=[(\tilde{l}_i, \tilde{e}_i)\mid 1\le i \le n]$ with $\tilde{l}_i=\{\tilde{p}^i_1, \ldots, \tilde{p}^i_m\}$, the attacked prediction result by LOA can be denoted as $g=[(l_i,e_i)\mid i\in[n] ]$ such that: $\forall i\in[n]$,
\[
l_i = \{\tilde{p}_1^i + (\beta, 0)\mid i\in[m]\}, \quad e_i = \tilde{e}_i,
\]
where $(\beta, 0)$ is the fixed horizontal offset applied to all lane points. 
An illustration attack is given in Figure~\ref{fig:visulize}(b).

\vspace{1mm}
\noindent\textbf{Lane Disappearance Attack.}  
LDA aims to remove all lane boundaries in an image, rendering the LD system inoperative. Under the LDA attack, the predicted result $g$ satisfies that: $\forall i\in[n]$,
\[
l_i = \varnothing, \quad e_i = 0.
\]
An illustration attack is given in Figure~\ref{fig:visulize}(c).

\vspace{1mm}
\noindent\textbf{Lane Rotation Attack.} For each lane boundary, a cubic spline interpolation is typically used to fit the curve. LRA aims to rotate such a curve around some starting point by a specified angle $\alpha$. The horizontal coordinates for the vertical coordinates in the label are then recalculated based on this rotation. 
Under the LRA attack, the predicted result $g$ satisfies that: $\exists i\in[n], j\in[m-1]$:
\[
l_i = \{\tilde{p}_1^i, \ldots, \tilde{p}_j^i, p_{j+1}^i, 
 \ldots, p_m^i\}, \quad e_i = \tilde{e}_i,
\]
where $\{p^i_{j+1},\ldots, p^i_m\}$ are rotated coordinates by some rotation angle $\alpha$. 
An illustration attack is given in Figure~\ref{fig:visulize}(d).

\vspace{1mm}
\noindent
\textbf{Backdoor Trigger.} We assume that the attacker aims to use practical physical triggers to mislead the LD models. These triggers could take the form of, for example, a cone with distinctive shapes or colors, or mud with a characteristic color following existing work~\cite{zhang2024towards,han2022physical}. 


\section{Methodology}

\subsection{Pineline of \tool}

As shown in Figure~\ref{fig:method_overview}, \tool consists of three stages: \textit{trigger position candidates generation}, \textit{optimal position selection}, and \textit{stealthy trigger generation}.

\noindent
{\bf Trigger Position Candidates Generation.} 
We segment the road region using SAM2~\cite{ravi2024sam2} to identify plausible areas for trigger placement. A sliding window is then applied over the segmented mask to generate a discrete set of position candidates.

\noindent
{\bf Optimal Trigger Position Finding.} Given the position candidates set, the optimal trigger position is then determined through a gradient-based heatmap analysis using a pre-trained LD network which shares the same architecture as the victim LD model. Moreover, different heatmap analysis approaches are designed to accommodate various attack strategies, ensuring optimal trigger placement in the most sensitive regions for each scenario.

\noindent
{\bf Trigger Generation.} Once the optimal trigger position is determined, a masked image is created by masking the trigger region, and the surrounding environmental context of the trigger position is extracted as the ground truth for the trigger's environment. This masked image is then fed into a carefully designed mask-guided diffusion model, which inpaints the trigger in a stealthy manner. To ensure seamless integration, the diffusion model is refined in the latent space through the incorporation of realness-preserving loss terms. These loss terms leverage the ground truth of the trigger's surrounding environment context to guide the generation process, ensuring that the synthesized trigger appears visually plausible and blends naturally into the scene.
Finally, the poisoned samples, containing the stealthily embedded triggers, are injected back into the training dataset. This would compromise the integrity of the LD model trained on the dataset, enabling the triggers to be effectively activated during the inference stage.

We next delve into the two core technical components that underpin the above stages in detail.


\subsection{Heatmap-based Optimal Trigger Position}\label{sec:heatmap}
 Prior studies \cite{wenger2021backdoor,abad2023sok} have demonstrated that embedding triggers in regions with higher attention enhances the ability of DNN models to memorize trigger patterns during training, thereby improving the attack's effectiveness during inference. However, conventional attention mechanisms, such as Grad-CAM \cite{selvaraju2017grad}, are predominantly tailored for pure classification tasks. In contrast, LD models inherently involve hybrid tasks, including \textit{lane existence} and \textit{lane coordinate} predictions, which require a more specialized approach. To address this gap and extract high-attention information relevant to our task, we propose a gradient-based attention map to identify optimal trigger positions. 

Formally, we design the gradient attention map \( M_{\text{grad}} \) to be computed using the following function:\vspace{-1mm}
\[
M_{\text{grad}}(i, j) = \sum_{k=1}^{CH} \left| \frac{\partial L}{\partial x_{i,j,k}} \right|, \text{for input image } x \in \mathbb{R}^{H \times W \times CH}
\]where \( L \) denotes the task-specific loss function, $(i, j)$ denotes the spatial position in the image, $k$ is the index of the channel dimension, and $CH$ represents the total number of channels. The resulting attention value $M_{\text{grad}}(i, j)$ is computed by summing the absolute gradient magnitudes across all channels at the given location, thereby capturing the sensitivity of that location with respect to the loss function. 

Moreover, to align with different attack strategies, we design \textit{strategy-specific heatmaps} by leveraging gradient information from task-relevant loss functions. Since different LD models are trained with different types of loss functions, we generate gradient attention heatmaps accordingly. Specifically, anchor-based methods (e.g., LaneATT~\cite{tabelini2021keep}, AdNet~\cite{xiao2023adnet}) typically use focal loss for lane existence prediction (used in LDA) and L1/GLIoU loss for coordinate regression (used in LOA and LRA). Segmentation-based methods (e.g., SCNN~\cite{pan2018spatial}, RESA~\cite{zheng2021resa}) rely on cross-entropy loss for classification and binary cross-entropy loss for regression, respectively. 
More details of these two types of LD are in supplementary materials.

\begin{figure*}[t]
    \centering
    \includegraphics[width=0.9\textwidth]{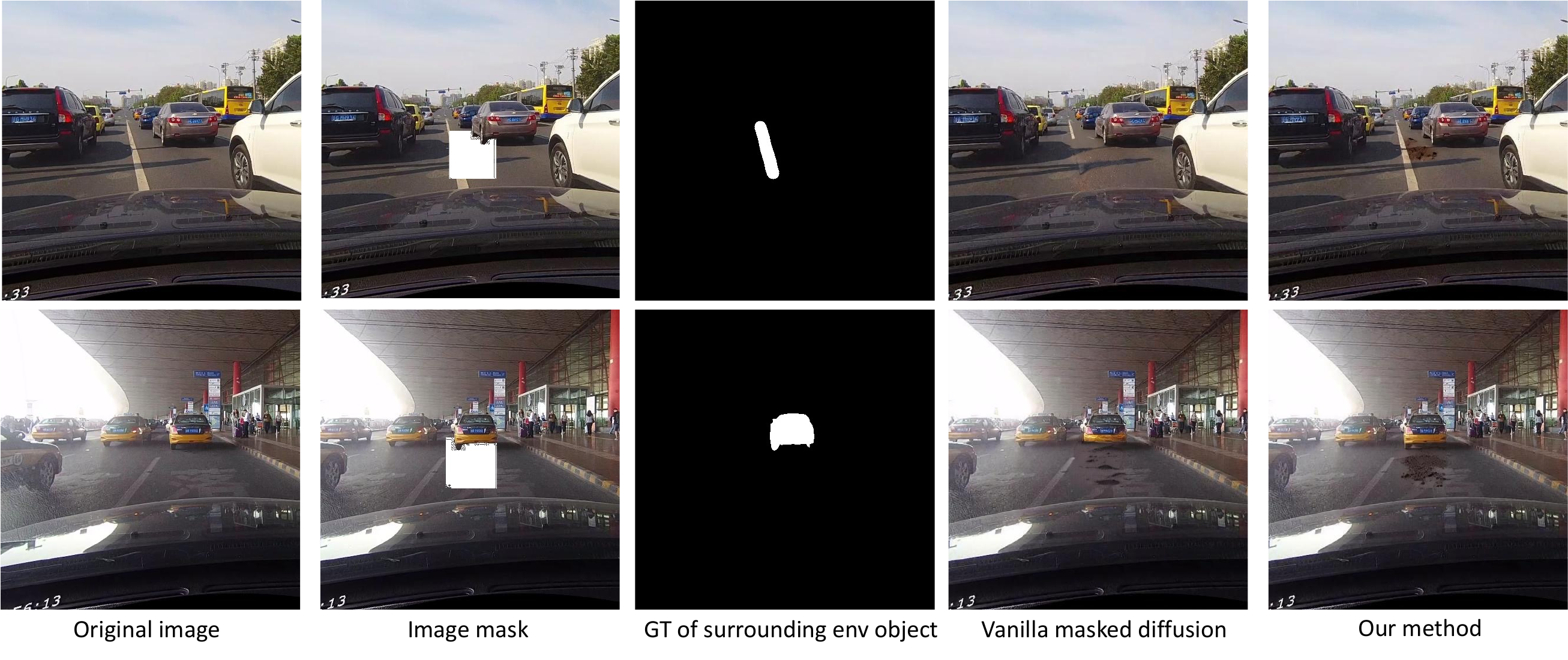}
    \caption{The bottom row demonstrates that, in the absence of the proposed lane and environment loss terms, the diffusion model may produce inconsistent
or unrealistic structures, such as deformed lane markings (above) and incomplete vehicles (below). In contrast, our method
effectively preserves contextual integrity, ensuring more accurate and visually coherent reconstructions.}
    \label{fig:vis}
\end{figure*}

Building on these foundations, we compute the optimal trigger position as follows. Given an input image and a specific attack strategy, we first generate a strategy-specific heatmap by leveraging the gradients of the corresponding loss functions. For example, in the case of LDA, classification loss gradients are used to suppress lane existence confidence, while for LOA or LRA, regression loss gradients are utilized to distort geometric features. Next, we identify high-sensitivity regions within the road areas segmented by {\sf SAM2}~\cite{ravi2024sam2}.
To pinpoint the most suitable locations for trigger placement, we employ a sliding window algorithm that identifies rectangular patches with maximal gradient responses within these high-sensitivity regions. This approach typically generates approximately $100\sim 400$ potential trigger candidates. By integrating the previously computed heatmap gradient results, we evaluate these candidates to determine the optimal trigger position with the highest attention value. The procedure ensures triggers are embedded in positions that exert the greatest influence on the targeted tasks, thereby maximizing the attack's efficacy.




    
    


\subsection{Diffusion-based Trigger Generator}

Based on the regions identified as optimal for trigger placement in Section~\ref{sec:heatmap}, we adopt a region-based diffusion editing framework inspired by {\sf UltraEdit} \cite{zhao2024ultraedit}, which facilitates precise and controlled injection of backdoor triggers within specific mask regions. Specifically, this mask-guided diffusion editing selectively modifies only the masked region, ensuring the generation of visually plausible and realistic backdoor triggers while leaving other areas of the image as unaffected as possible. 

Let $z_t \in \mathbb{R}^{H \times W \times C}$ denote the latent representation of the image at diffusion step $t$, and $z_T$ be the initial noisy latent sampled from a standard Gaussian distribution. The overall diffusion update process in {\sf UltraEdit} can then be formulated as follows:
\begin{equation*}
z_{t-1} = 
\begin{cases}
(1 - M) \odot z_T + M \odot D_M(z_t), & \text{If } t \bmod 2 = 0, \\
D_M(z_t), & \text{Otherwise},
\end{cases}
\end{equation*}
where $\odot$ denotes element-wise multiplication, and $M \in \{0,1\}^{H \times W \times C}$ is the binary mask that specifies the trigger region,  $M(i,j,c) = 1$ demonstrates that the pixel at position $(i,j)$ in channel $c$ is designated for modification.  $D_M(z_t)$ refers to the denoising function applied to $z_t$, with updates restricted to the masked region defined by $M$. When the diffusion step is even, the diffusion process is confined strictly to the specified masked region, leaving the latent encoding of the original image unchanged outside the mask. Conversely, when the diffusion step is odd, standard diffusion is applied across the entire image. Due to the preservation of the original latent encoding during the even steps, regions outside the mask remain largely unaffected.

Through this alternating procedure, the mask region undergoes precise modification, while the visual integrity and consistency of regions outside the mask are effectively preserved. We argue that this mechanism is particularly well-suited for generating backdoor triggers at specific locations, as it ensures stealthiness by minimizing unintended modifications to other parts of the image.

However, the inherent attention mechanisms of diffusion models still inadvertently distort crucial visual elements, such as lane markings or surrounding traffic objects located adjacent to the mask boundaries. To mitigate this issue and improve the stealthiness of the attack, we introduce two additional loss terms, $L_{\text{lane}}$ and $L_{\text{env}}$, into the diffusion process:

\vspace{1mm}
\scalebox{0.9}{$
\begin{aligned}
    L_{\text{lane}} &= \text{MSE} \left( \text{gen\_img} \odot \text{lane\_mask},\ \text{clean\_img} \odot \text{lane\_mask} \right) \\
    L_{\text{env}}  &= \text{SSIM} \left( \text{gen\_img} \odot \text{env\_mask},\ \text{clean\_img} \odot \text{env\_mask} \right)
\end{aligned}
$}

\vspace{1mm}

\noindent $L_{\text{lane}}$ is designed to guide the generated image to closely align with the original (clean) image within the lane regions, as specified by the binary \text{lane\_mask}. By minimizing this mean squared error, we ensure that the lane structures remain visually intact and consistent, thereby reducing potential artifacts that could otherwise compromise the realism of the scene. $L_{\text{env}}$, on the other hand, employs structural similarity (SSIM) to preserve the visual integrity and realism of surrounding traffic elements and environmental structures. This ensures that the modifications made during the diffusion process do not distort the contextual coherence of the scene, maintaining a high level of stealthiness in the attack.




Incorporating these two loss terms is essential for improving the stealthiness of the generated triggers, as they effectively suppress the artifacts typically introduced by diffusion models. This design choice enables the creation of poisoned samples that appear seamless and natural. As illustrated in Figure~\ref{fig:vis}, removing these loss terms (i.e., adopting vanilla mask-guided diffusion) often leads to noticeable inconsistencies in the reconstructed content, especially in areas adjacent to the mask. In contrast, our proposed method preserves critical environmental cues and yields reconstructions that remain visually coherent and semantically faithful to the original scene.

\paragraph{Clarification on Trigger Diversity.} To quantify trigger diversity, we compute pairwise perceptual similarity using LPIPS (Learned Perceptual Image Patch Similarity) scores~\cite{zhang2018unreasonable} and adopt a threshold of 0.15, following prior work~\cite{hou2022perceptual}. This threshold guides the construction of practical triggers by ensuring that each trigger retains a distinctive visual signature while remaining sufficiently dissimilar from benign patterns that may, albeit rarely, appear in the dataset. Consequently, it preserves the clean-test accuracy of the poisoned model while maintaining a high ASR. 

\begin{table*}[t]
\centering
\renewcommand{\arraystretch}{0.9}
\setlength{\tabcolsep}{3pt}
\small
\scalebox{0.8}{
\begin{tabular}{c|c|c|c|ccc|c|c|c|cc|c}
\toprule
\multirow{4}{*}{\textbf{Model}}&\multirow{4}{*}{\textbf{Attack}}&\multicolumn{6}{c|}{\textbf{CULane}}&\multicolumn{5}{c}{\textbf{TuSimple}}\\
\cmidrule(lr){3-8}\cmidrule(lr){9-13}
&&\multicolumn{2}{c|}{\textbf{ACC on clean images}}&\multicolumn{3}{c|}{\textbf{ASR (Diff. strategies)}}&\textbf{ASR}&\multicolumn{2}{c|}{\textbf{ACC on clean images}}&\multicolumn{2}{c|}{\textbf{ASR (Diff. strategies)}}&\textbf{ASR}\\
\cmidrule(lr){3-4}\cmidrule(lr){5-7}\cmidrule(lr){8-8}\cmidrule(lr){9-10}\cmidrule(lr){11-12}\cmidrule(lr){13-13}
&&Benign& Poisoned &LDA&\multicolumn{1}{c}{\hspace{1em}LRA\hspace{1em}}&\hspace{1em}LOA&\textbf{Avg.}& Benign & Poisoned&\hspace{0.8em}LDA&\hspace{0.8em}LOA&\textbf{Avg.}\\
\midrule
\multirow{6}{*}{\sf LaneATT}& {\sf BadNets}&~ &72.79&49.27&\hspace{0.6em}43.37&\hspace{0.6em}47.36&46.67&~&95.53&\hspace{0.8em}45.62&\hspace{0.8em}75.63& 60.63\\
&{\sf Blended}&&73.15&57.14&\hspace{0.6em}59.07&\hspace{0.6em}47.04&54.41&&96.27&\hspace{0.8em}46.31&\hspace{0.8em}72.47&59.39\\
&{\sf LD-Attack}&
{75.01}&72.48&71.24&\hspace{0.6em}56.24&\hspace{0.6em}63.71&63.73&
{96.81}&96.21&\hspace{0.8em}88.17&\hspace{0.8em}90.27&89.22\\
&{\sf BadLANE}&&73.78&64.33&\hspace{0.6em}45.22&\hspace{0.6em}54.15&54.56&&96.18&\hspace{0.8em}89.25&\hspace{0.8em}89.76&89.51\\
&\textbf{\tool}&&74.20&\textbf{81.65}&\hspace{0.72em}\textbf{73.45}&\hspace{0.65em}\textbf{74.19}&\textbf{76.41}&&95.72&\hspace{0.8em}\textbf{90.04}&\hspace{0.8em}\textbf{91.63}&\textbf{90.84}\\
\midrule
\multirow{6}{*}{{\sf ADNet}}&{\sf BadNets}&\multirow{6}{*}[0.55em]{85.51}&79.60&50.13&\hspace{0.6em}54.33&\hspace{0.6em}61.39&55.28&\multirow{6}{*}[0.55em]{96.87 }&95.46&\hspace{0.8em}50.34&\hspace{0.8em}52.98&51.66\\
&{\sf Blended}&&74.37&43.67&\hspace{0.6em}56.42&\hspace{0.6em}56.51&52.20&&95.10&\hspace{0.8em}56.27&\hspace{0.8em}59.12&57.70\\
&{\sf LD-Attack}&&79.14&59.47&\hspace{0.6em}63.39&\hspace{0.6em}67.15&63.34&&96.35&\hspace{0.8em}53.24&\hspace{0.8em}66.32&59.78\\
&{\sf BadLANE}&&85.46&36.61&\hspace{0.6em}61.16&\hspace{0.6em}65.95&54.57&&96.29&\hspace{0.8em}59.17&\hspace{0.8em}67.48&63.33\\
&\textbf{\tool}&&82.37&\textbf{66.49}&\hspace{0.6em}\textbf{64.82}&\hspace{0.6em}\textbf{68.17}&\textbf{66.50}&&96.61&\hspace{0.8em}\textbf{89.24}&\hspace{0.8em}\textbf{72.07} &\textbf{80.66 }\\
\midrule
\multirow{6}{*}{{\sf SCNN}}&{\sf BadNets}&\multirow{6}{*}[0.55em]{69.75}&69.52&74.39&\hspace{0.6em}63.63&\hspace{0.6em}62.76&66.93&\multirow{6}{*}[0.55em]{93.78}&91.56&\hspace{0.8em}64.72&\hspace{0.8em}62.79&63.76\\
&{\sf Blended}&&63.98&75.14&\hspace{0.6em}63.45&\hspace{0.6em}61.60&66.73&&91.37&\hspace{0.8em}58.30&\hspace{0.8em}64.13&61.22\\
&{\sf LD-Attack}&&64.38&74.83&\hspace{0.6em}62.21&\hspace{0.6em}64.17&67.07&&92.23&\hspace{0.8em}94.08&\hspace{0.8em}\textbf{91.43}&92.76\\
&{\sf BadLane}&&67.57&72.28&\hspace{0.6em}63.96&\hspace{0.6em}62.94&66.39&&90.89&\hspace{0.8em}92.19&\hspace{0.8em}88.57&90.38\\
&\textbf{\tool}&&68.75&\textbf{76.19}&\hspace{0.6em}\textbf{67.34}&\hspace{0.6em}\textbf{68.69}&\textbf{70.74}&&92.28&\hspace{0.8em}\textbf{95.78}&\hspace{0.8em}91.29&\textbf{93.54}\\
\midrule
\multirow{6}{*}{{\sf RESA}}&{\sf BadNets}&\multirow{6}{*}[0.55em]{78.98}&76.68&71.08&\hspace{0.6em}58.68&\hspace{0.6em}69.14&66.30&\multirow{6}{*}[0.55em]{96.81}&96.63&\hspace{0.8em}64.78&\hspace{0.8em}73.46&69.12\\
&{\sf Blended}&&75.87&73.85&\hspace{0.6em}59.64&\hspace{0.6em}57.76&63.75&&96.65&\hspace{0.8em}68.11&\hspace{0.8em}57.28&62.70\\
&{\sf LD-Attack}&&71.35&73.26&\hspace{0.6em}64.77&\hspace{0.6em}69.18&69.07&&96.67&\hspace{0.8em}92.53&\hspace{0.8em}92.04&92.29\\
&{\sf BadLANE}&&72.81&75.39&\hspace{0.6em}62.94&\hspace{0.6em}68.23&68.85&&96.42&\hspace{0.8em}92.47&\hspace{0.8em}91.19&91.83\\
&\textbf{\tool}&&77.39&\textbf{78.15}&\hspace{0.6em}\textbf{65.15}&\hspace{0.6em}\textbf{76.86}&\textbf{73.39}&&96.67&\hspace{0.8em}\textbf{96.57}&\hspace{0.8em}\textbf{94.28}&\textbf{95.43}\\
\bottomrule
\end{tabular}}
\caption{ACC and ASR(\%) of different attack methods on the CULane and TuSimple datasets under various attack strategies.}
\label{tab:culane_tusimple_asr}
\end{table*}

\section{Evaluation}\label{sec:exp}
%

    
    
    

\subsection{Experiment Setup}

\textbf{Datasets.} We use CULane~\cite{pan2018SCNN} and TuSimple~\cite{pizzati2019lane}, containing $88$K/$34$K and $3.6$K/$2.7$K train/test images respectively, with resolutions $1,640{\times}590$ and $1,280{\times}720$, and up to 4 or 5 lanes per frame.

\noindent
\textbf{Lane Detectors.} We evaluate four representative models: anchor-based model (LaneATT~\cite{tabelini2021keep}, ADNet~\cite{xiao2023adnet}) and segmentation-based model (SCNN~\cite{pan2018spatial}  RESA~\cite{zheng2021resa}).

\vspace{1mm}\noindent
\textbf{Attack Baseline.} Considering practical deployment in autonomous driving and LD tasks, we adopt four state-of-the-art and physically realizable backdoor attack strategies as our baselines:
\textbf{\textcircled{1} Fixed Pattern - {\sf BadNets}~\cite{gu2019badnets}:} Adds a fixed white square at the bottom-right corner of the image as a static trigger pattern.
\textbf{\textcircled{2} Fixed Image - {\sf Blended}~\cite{chen2017targeted}:} Blends a predefined universal image trigger with the input sample at a fixed blending ratio.
\textbf{\textcircled{3} Fixed Pattern \& Image - {\sf LD-Attack \cite{han2022physical}}:} Uses a fixed trigger overlay and blends it with the input image.
\textbf{\textcircled{4} Random Texture - {\sf BadLane} \cite{zhang2024towards}:} Introduces randomly placed mud-like texture noise across the road surface, simulating naturally occurring triggers.

\noindent
\textbf{Evaluation Metric.} 
We follow standard TuSimple protocol~\cite{zhang2024towards}, reporting Accuracy (ACC) on clean data and Attack Success Rate (ASR) under LOA, LRA, and LDA. ACC and ASR for LOA/LRA are defined as $\text{Metric} = \frac{\sum_i C_i}{\sum_i S_i}$, where $C_i$ is the number of correct lane points (within 20px of GT), and $S_i$ is the total. For LDA, ASR is computed as the fraction of disappeared lanes.

\noindent
\textbf{Implementation.}
We set the poisoning rate to 10\%, with 900-pixel triggers. For {\sf BadNets}/{\sf Blended}, a $30{\times}30$ square is placed bottom-right.  {\sf LD-Attack} place  900-pixel on the adjacent lane near the boundary. {\sf BadLane} samples 900 pixels from a $100{\times}100$ region. Our method uses a $40{\times}40$ mask with diffusion-based trigger generation, using prompts such as ``Add a small and sparse brown mud'' or ``Add a small traffic cone''. UltraEdit~\cite{zhao2024ultraedit} is used for generation.
For attack parameters, we follow prior work and set the LOA offset to 60 pixels and the LRA rotation to 9 degrees.

\subsection{The Effectiveness of \tool}\label{sec:rq1}

We compare different backdoor attack baselines on four LD models across the CULane and TuSimple datasets. Specifically, we evaluate: (1) the accuracy (ACC) of original/benign and poisoned LD models on the clean images to assess the clean accuracy drop of different attack methods; and (2) the ASR under various backdoor injection strategies, including LDA, LRA, and LOA.

The experimental results tested on the mud trigger are given in Table~\ref{tab:culane_tusimple_asr}, from which we draw the following key observations:
{\bf i) CULane is more challenging to attack than TuSimple.} 
The CULane dataset consists of complex urban driving scenes with dynamic objects such as pedestrians, traffic lights, and intersections, making it significantly more difficult to manipulate compared to the highway-based TuSimple dataset. Traditional attack methods perform poorly on both datasets, and although {\sf Badlane} achieves relatively high ASR on TuSimple (e.g., 90.84\% on LaneATT), its performance drops significantly on CULane (only 54.56\%). We remark that this highlights \emph{the difficulty of launching successful physical backdoor attacks in more diverse and cluttered environments};
\textbf{ii) \tool achieves the highest ASR across all attack strategies.}
Our proposed method almost outperforms all baseline methods under LDA, LRA, and LOA. Among these, LDA consistently yields the highest ASR, likely due to its simplicity in altering lane direction. In contrast, LOA and LRA are harder to execute but pose greater safety risks by shifting the vehicle trajectory. LOA achieves higher ASR than LRA, which may be attributed to a broader label shift. Overall, \tool improves the average ASR by a substantial margin and generalizes well across different lane detection models. This further demonstrates that the physical trigger is more effective in activating the backdoor. 
\textbf{iii) \tool maintains high accuracy on clean samples.}  Despite its strong attack capability, \tool introduces only a minor degradation in model performance on clean inputs. Poisoned models retain comparable accuracy to their vanilla counterparts, suggesting that the injected backdoor remains stealthy and does not interfere with normal behavior.
\textbf{iv) ADNet is more robust than other lane detection models.} 
LaneATT, RESA, and SCNN models exhibit high ASRs under \tool, nearly matching their clean ACCs, indicating high vulnerability. In contrast, ADNet shows greater resilience, with a notable gap of over 16\% between its ASR and ACC, suggesting ADNet plays a role in robustness against physical backdoor attacks. 
More results, e.g., ASR of other triggers and computation resources, are provided in the supplementary. 

\subsection{The Stealthiness of \tool}
To evaluate the stealthiness of injected backdoors, we adopt three representative image-based forensic detection methods due to the lack of LD-specific backdoor detectors. \textbf{UniDetection (U)}~\cite{ojha2023towards} uses CLIP-based features to identify subtle inconsistencies in generated content. \textbf{LGrad (L)}\cite{tan2023learning} computes saliency maps from CNN feature gradients to highlight local artifacts. \textbf{DIRE (D)}~\cite{wang2023dire} is tailored for detecting diffusion-generated images by modeling visual trace patterns.

We feed both benign and poisoned samples into each detector. All three methods achieve over 90\% 
accuracy on benign samples. The results for poisoned samples are summarized in Table~\ref{tab:backdoor_methods}, where \textbf{N/A} denotes direct patching methods that require no forensic analysis and are easily detectable by human inspection.
{\sf BadNets}, {\sf Blended} and  {\sf LD-Attack}  remain largely undetected, likely due to their trigger being directly overlaid onto the image without introducing detectable artifacts. In contrast, UniDetection detects nearly 60\% of {\sf Badlane}'s backdoored samples, while only identifying less than 3\% of \tool's samples. Unfortunately, LGrad fails to effectively detect both \tool and {\sf Badlane}, possibly because the detector is not fine-tuned on autonomous driving related scenarios. Additionally, DIRE proves ineffective against \tool samples. This is expected, as \tool only modifies a small region of the image while preserving most of the original content. 
Note that, although {\sf BadNets}, {\sf Blended} and {\sf LD-Attack} exhibit lower detection rates, they are visually conspicuous and thus vulnerable to manual inspection. while the mud trigger from \tool is visually natural and blends into the road context, making it much harder to identify through manual observation. 
More trigger visualizations are available in the \cite{dbald_site}.

\begin{table}[t]
\centering
\renewcommand{\arraystretch}{1.2}
\setlength{\tabcolsep}{7pt}
\setlength{\tabcolsep}{3pt}
\scalebox{0.85}{\begin{tabular}{l|c|c|c|c|c|c}
 \toprule
 \multirow{2}{*}{\textbf{Method}} & \multicolumn{3}{c|}{\textbf{CULane}} & \multicolumn{3}{c}{\textbf{TuSimple}} \\
 \cmidrule(lr){2-4} \cmidrule(lr){5-7}
 & \textbf{U} & \textbf{L} & \textbf{D} & \textbf{U} & \textbf{L} & \textbf{D} \\
 \hline
 {\sf BadNets} & \textbf{2.56} & N/A & N/A & \textbf{0.08} & N/A & N/A \\
 \hline
 {\sf Blended} & 5.18 & N/A & N/A & 0.17 & N/A & N/A \\
 \hline
 {\sf LD-Attack} & 6.27 & N/A & N/A & 1.31 & N/A & N/A \\
 \hline
 {\sf Badlane} & 59.31 & 9.17 & N/A & 35.60 & 4.23 & N/A \\
 \hline
 \textbf{\tool} & 2.72 & \textbf{1.76} & 0.56 & 0.23 & 0.51 & 0.04 \\
 \bottomrule
\end{tabular}}
\caption{Detection accuracy (\%) of various backdoor attack methods under different forensic detectors.}
\label{tab:backdoor_methods}
\end{table}

\subsection{Ablation Study}
\textbf{Effectiveness of Heatmap Guidance.}  
We investigate the contribution of heatmap guidance in enhancing the attack robustness of \tool. All experiments are conducted under the \textit{mud} trigger on the CULane dataset. 
Results are given in Table~\ref{tab:lane_method_comparison}, where ``Without heatmap'' means that the mud trigger is placed randomly during poisoning without any heatmap-based guidance. We find that using heatmap guidance significantly improves ASR on both models: \textbf{+3.21\%} on LaneATT and \textbf{+6.77\%} on RESA. These results demonstrate that spatially guided trigger placement enhances the robustness of the backdoor attack. 

\noindent\textbf{Effectiveness of Diffusion Losses.}
We further perform ablation on the two proposed diffusion losses, $L_{\text{lane}}$ and $L_{\text{env}}$. Specifically, under forensic detection~\cite{ojha2023towards}, the detection scores increase from \textbf{2.46} (with both losses) to \textbf{12.13} (w/o lane loss) and \textbf{14.51} (w/o environment loss). These results highlight that both loss terms are essential in producing perceptually natural poisoned images that can better evade existing detection techniques.

\begin{table}[t]
\centering
\renewcommand{\arraystretch}{0.9}
\scalebox{0.8}{\begin{tabular}{l|ccc|ccc}
\toprule
\multirow{2}{*}{\textbf{Method}} & \multicolumn{3}{c|}{\textbf{LaneATT}} & \multicolumn{3}{c}{\textbf{RESA}} \\
\cmidrule(lr){2-4}\cmidrule(lr){5-7}
& LDA & LRA & LOA & LDA & LRA & LOA \\
\midrule
With heatmap & 81.65 & 73.45 & 74.19 & 78.15 & 65.15 & 76.86 \\
Without heatmap & 80.97 & 70.16 & 68.52 & 74.23 & 58.26 & 69.37 \\
\bottomrule
\end{tabular}}
\caption{Attack performance w/wo the heatmap guidance.}
\label{tab:lane_method_comparison}
\end{table}


%
\subsection{Physical Experiment}
To better simulate real-world conditions, we go beyond prior sandbox setups~\cite{zhang2024towards} and conduct physical tests in four closed-road environments. We use two types of cones with different shapes and colors to simulate diverse trigger appearances and evaluate attack performance on the poisoned LaneATT model under the LOA setting.

\begin{figure}[t]
    \centering
    \includegraphics[width=0.9\linewidth]{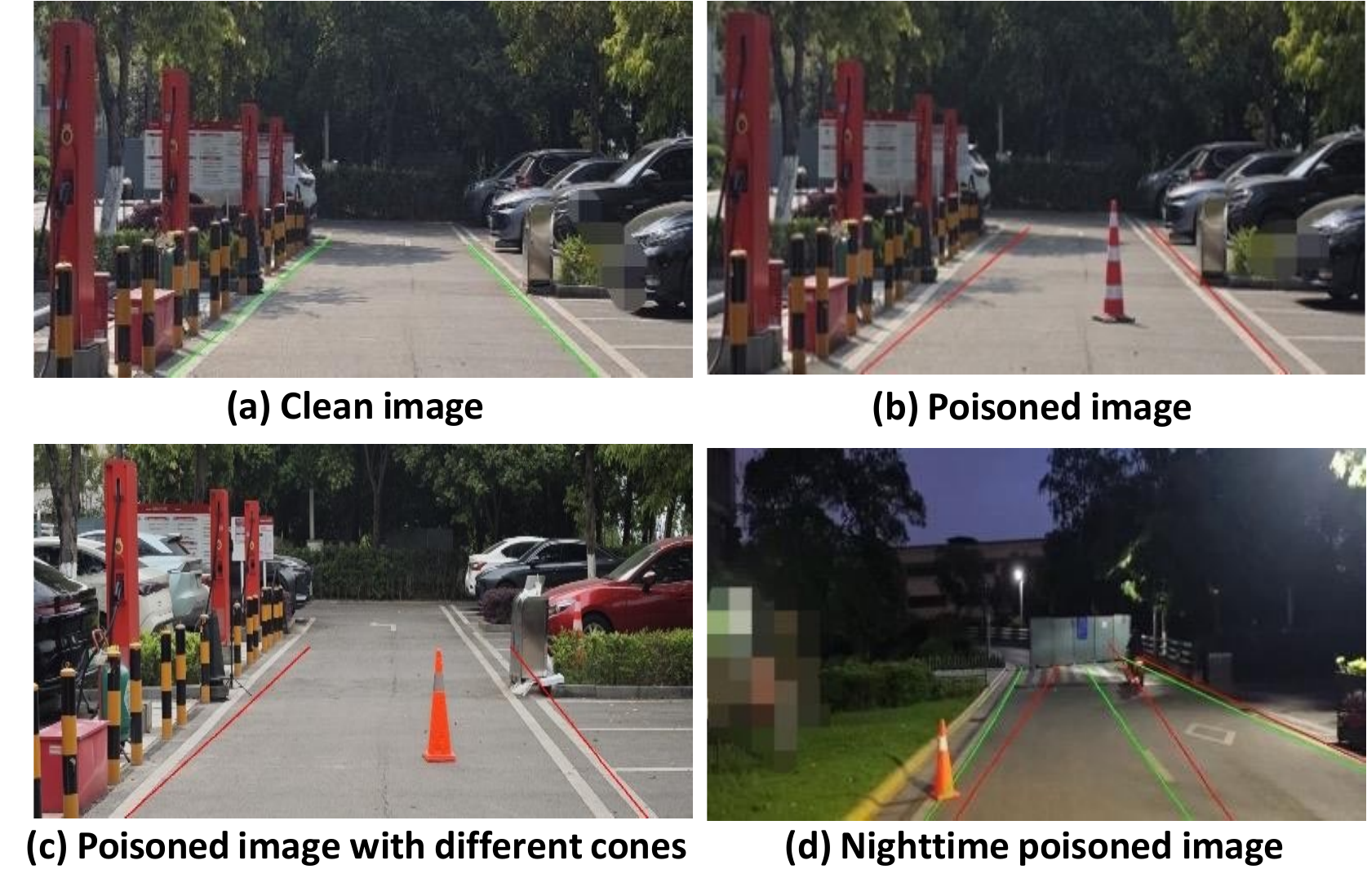}
    \vspace{-1 mm}
    \caption{Physical attacks under different conditions using cone triggers.}
    \label{fig:psysical_trigger}
\end{figure}

We collect 100 poisoned images using cones across both daytime and nighttime conditions with diverse environments and placements, yielding a 57\% ASR. The findings suggest that \tool remains potentially effective under varying conditions in physical scenarios, and thus poses a real threat to LD models in the physical world. Note that, although lower than simulation-based results, it is expected as the model is not retrained on the physical domain. Additionally, using SOTA weather transfer methods~\cite{batifol2025flux,dong2023benchmarking}, we apply rain, blur, and occlusion to 100 poisoned images, observing only slight ASR drops (from 57\% to 45/48/49\%), indicating our method's robustness to diverse environments. 
Figure~\ref{fig:psysical_trigger} shows benign and attack cases across different cones and lighting conditions. 

\section{Countermeasures}


We evaluate several defense strategies against \tool, following prior works~\cite{han2022physical, zhang2024towards}. For fine-tuning, we retrain the poisoned LaneATT model (under LOA) on clean data for 50 epochs, which reduces ASR from 74.19\% to 63.68\%, showing partial but insufficient mitigation. For neuron pruning, we progressively remove neurons from the final convolutional layer (512 in total) in steps of 50. While pruning 50 neurons only slightly reduces ASR (to 73.67\%), more aggressive pruning (e.g., 400 neurons) lowers ASR to 15.87\%, but also severely degrades clean accuracy (ACC drops from 74.20\% to 21.19\%). This suggests pruning is largely ineffective unless it sacrifices benign performance.
We also assess a vision-language model Qwen2.5-VL\cite{bai2025qwen3} as an anomaly detector. Sampling 100 poisoned images, we use the model to identify visual anomalies. Detection rates are low: only 6\% for cone-based triggers and 3\% for mud-based triggers, indicating that multimodal models struggle to detect our triggers. 
We will investigate the how the quantization and reasoning injection make effect in mitigating the attack for future work~\cite{liu2024quantized}. 

\section{Conclusion and Future Work}
We propose \tool, a diffusion-based backdoor framework for LD systems that places triggers in sensitive regions guided by gradient attention. It supports multiple attack strategies via strategy-specific heatmaps and improves stealthiness through lane preservation and environment consistency losses. Experiments show \tool outperforms prior backdoors in both digital and physical settings.
For future improvements, we believe a task-specific distilled diffusion model could enhance efficiency. While we currently use UltraEdit as a general-purpose backbone, a tailored model trained to seamlessly embed triggers in driving scenes may substantially reduce diffusion steps and costs.


\bibliographystyle{unsrt}
\bibliography{main} 

\appendix
\section{Diffusion Denoising}
Diffusion-based inpainting methods extend standard diffusion models by introducing spatial constraints, where a mask specifies the region to be edited while preserving the remaining content. 
Concretely, the model starts from a noisy latent representation and iteratively denoises it, but unlike unconditional generation, the denoising process is conditioned not only on textual instructions but also on the input image and a binary or soft mask that restricts modifications to a target region. 
In UltraEdit~\cite{zhao2024ultraedit}, this is implemented via a modified inpainting diffusion pipeline that alternates between global denoising and masked updates, ensuring that edits are localized while maintaining overall image consistency. 
Building upon this framework, the incorporation of additional losses can be naturally integrated into the denoising process.
Therefore, to maintain reality, after decoding the latent into image space two structure-preserving losses constrain the edited region to remain coherent with the original scene. 
These losses back-propagate through the denoising steps, effectively guiding the noise prediction toward solutions that are both instruction-faithful and visually plausible.

\section{More Experiment results}
\subsection{ASR under varying environments}
To evaluate the performance of \tool under varying physical triggers and driving environments, we test the LOA attack using mud and cone triggers across various lightness conditions in the CULane dataset. Results are given in Table~\ref{tab:asr_env_trigger}.
We find that both mud and cone triggers achieve relatively high ASRs across all environments, indicating their effectiveness under LOA. Notably, \textbf{the cone trigger outperforms mud at night across all models, with an improvement of up to 5\%}, which is reasonable since cones are more visually distinguishable than mud in low-light conditions, resulting in more reliable backdoor activations.

We also find that clean accuracy drops in challenging conditions, which may affect backdoor sensitivity. For instance, under the LOA setting with RESA, clean accuracy decreases from 84.77\% (normal) to 82.05\% (shadow) and 77.37\% (night). This supports our hypothesis that low-light conditions degrade LD model performance, which could inadvertently make the model less reactive to certain perturbations like mud triggers, while still allowing salient triggers (e.g., cones) to remain effective.

\begin{table}[h]
\caption{ASR(\%) for different LD models with varying triggers and environments on the CULane dataset.}
\setlength{\tabcolsep}{3pt}
\scalebox{0.85}{
\begin{tabular}{l|l|c|c|c|c}
\midrule
\multirow{2}{*}{\textbf{Trigger}} & \multirow{2}{*}{\textbf{Model}} 
& \multicolumn{4}{c}{\textbf{LOA ASR (Varying Driving Environments)}} \\
\cmidrule{3-6}
& & Normal & Shadow & Highlight & Night \\
\midrule

\multirow{4}{*}{Mud} 
& LaneATT & 74.54 & 72.05 & 74.21 & 67.42 \\
& ADNet   & 69.51 & 67.34 & 67.17 & 62.23 \\
& SCNN    & 67.54 & 68.64  & 68.51 & 62.46 \\
& RESA    & 77.06 & 75.11 & 77.24 & 72.26 \\
\midrule

\multirow{4}{*}{Cone} 
& LaneATT & 74.17 & 69.95 & 73.30 & 70.13 \\
& ADNet   & 68.62 & 68.58 & 64.03 & 65.47 \\
& SCNN    & 68.93 & 69.17 & 69.24 & 66.32 \\
& RESA    & 77.01 & 76.34 & 73.29 & 75.17 \\
\midrule

\end{tabular}}
\label{tab:asr_env_trigger}
\end{table}

\subsection{ASR under different attack parameters}
\begin{figure}[t]
  \centering
  \includegraphics[width=0.7\linewidth]{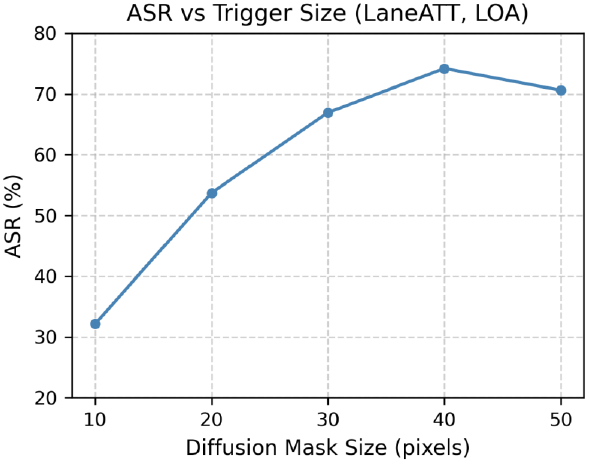}
  \caption{ASR vs. Trigger Size for mud trigger under the LOA setting (LaneATT, CULane).}
  \label{fig:asr_vs_trigger_size}
\end{figure}

We further explore the impact of trigger size under the LOA setting using the LaneATT model and CULane dataset with mud triggers. As shown in Figure~\ref{fig:asr_vs_trigger_size}, smaller triggers yield significantly lower attack success rates (ASR), with ASR dropping from 74.19\% (40$\times$40) to 32.15\% (10$\times$10). This suggests that tiny triggers are less effective at activating the backdoor. However, ASR slightly drops when the trigger size exceeds $40{\times}40$ (e.g., 70.62\% at $50{\times}50$). This is due to a corresponding drop in clean accuracy—excessively large triggers distort the input scene, degrading overall model performance and thus suppressing backdoor activation. Notably, even with a trigger as small as 20$\times$20 pixels, the ASR exceeds 50\%, indicating that DBALD remains effective under compact trigger settings. This supports its potential stealthiness in practical deployment.

\subsection{ASR under different poisoned ratios}
We further study how different poisoning rates affect the attack performance of \tool. On both LaneATT and RESA models, we train LD models using poisoning rates of \textbf{1\%, 3\%, 5\%, 10\%} and \textbf{15\%} on the CULane dataset. For reference, constructing a 1\% poisoned dataset (i.e., 177 poisoned samples) takes approximately 3 minutes. The results are given in Table~\ref{tab:poisoning_rate}. We find that \tool achieves relative strong attack performance even at very low poisoning rates. As the poisoning rate rises, the ASR improves gradually, while the accuracy on clean samples decreases slowly.

\begin{table}[h]
\centering
\setlength{\tabcolsep}{8pt}
\caption{Attack performance at varied poisoning rates.}
\scalebox{0.85}{\begin{tabular}{l|cc|cc}
\toprule
\multirow{2}{*}{Poisoning Rate} & \multicolumn{2}{c|}{LaneATT} & \multicolumn{2}{c}{RESA} \\
\cmidrule(lr){2-3}\cmidrule(lr){4-5}
 & ACC & ASR & ACC & ASR \\
\midrule
\hspace{2em}1\% & \textbf{75.26} & 64.27 & \textbf{79.98} & 68.23 \\
\hspace{2em}3\% & 75.10 & 72.24 & 79.18 & 74.87 \\
\hspace{2em}5\% & 75.14 & 73.52 & 79.12 & 75.46 \\
\hspace{2em}10\% &  74.20 &  \textbf{74.19} & 77.39 & \textbf{76.86} \\
\hspace{2em}15\% & 73.23 & 73.10 & 77.20 & 76.48 \\
\bottomrule
\end{tabular}}
\label{tab:poisoning_rate}
\end{table}

\subsection{Justification of the attention mechanism}

The proposed gradient-based attention mechanism is grounded in both prior work and our analytical findings.

Prior work~\cite{wenger2021backdoor} demonstrates that gradient-based high-attention regions are strongly correlated with high ASR, suggesting that effective triggers must be strategically placed. Motivated by this, we introduce a similar gradient-driven strategy tailored to lane detection (LD) tasks.

We further conduct an analytical study by visualizing the gradient-based attention maps over training epochs for both clean and trigger-injected images. We observe that:

\begin{itemize}
  \item For clean images, attention gradually concentrates on task-relevant semantic regions.
  \item For trigger-injected images, attention becomes increasingly focused on the trigger region across training.
\end{itemize}

This evolving focus indicates that the model gradually ``memorizes'' the trigger position, reinforcing the effectiveness of targeted trigger placement. Table~\ref{tab:attn_analysis} summarizes results at epoch 100.

\begin{table}[h]
\centering
\caption{Comparison of gradient-based attention maps at epoch 100.}
\label{tab:attn_analysis}
\scalebox{0.9}{
\begin{tabular}{l|c|c}
\toprule
\textbf{Setting} & \textbf{Attention Entropy $\downarrow$} & \makecell{\textbf{Attention on}\\\textbf{Trigger Region $\uparrow$}} \\
\midrule
Trigger (Ours)    & 16.09 & 47.39\% \\
Random Trigger     & 16.26 & 44.13\% \\
Clean Image        & 16.96 & 41.58\% \\
\bottomrule
\end{tabular}
}
\end{table}

The lower entropy and higher focus on the trigger region for our method suggest that the model allocates more stable and intense attention to the trigger, thereby improving attack reliability.

\subsection{Computation time}
\tool incurs a moderate computational cost relative to baseline methods (6 seconds per image vs. 1 second). Since data poisoning is a one-time offline procedure, this computational overhead is generally acceptable. 

\subsection{Clarification on trigger diversity}

To preserve high clean accuracy and maintain a low false-positive (FP) rate—particularly in the rare cases where benign patterns resemble potential triggers—we use LPIPS scores to distinguish our triggers from benign visual patterns in the original data. We adopt a threshold of 0.15 to differentiate benign patterns from actual triggers. On benign-only datasets (e.g., TuSimple), this yields a false-positive rate as low as 8.2\%.
It is important to note that: i) false positives are inevitable in any backdoor attack when considered in an open-world usage setting; and ii) backdoor triggers can be non-fixed and may even be designed to vary dynamically during inference to achieve high ASR. 

In this work, our priority is to achieve high ASRs with realistic, context-aware triggers while keeping the false-positive rate low. To this end, we maintain two lightweight reference databases throughout training and evaluation:

\begin{itemize}[leftmargin=12pt]
    \item \textbf{Benign pattern database}: a collection of naturally occurring benign visual patterns in clean images that are semantically similar to potential triggers.
    \item \textbf{Poisoned trigger database}: a repository of all synthesized triggers used during the backdoor injection process.
\end{itemize}
These two databases ensure that each newly synthesized trigger remains sufficiently distinct from benign patterns (LPIPS $>0.15$ when compared against the benign pattern database) while remaining visually consistent with existing poisoned triggers (LPIPS $< 0.15$ when compared against the poisoned trigger database), thereby preserving clean accuracy, maintaining low false-positive rates, and sustaining high ASR with context-aware triggers.

We note that although the clean dataset is large, benign trigger-like patterns are extremely rare (e.g., mud-like textures do not appear, and cone-like shapes occur in only 129 images), resulting in a very small benign-pattern database; the poison-trigger database is similarly small, containing only about 1\% of the data corresponding solely to injected trigger instances. Owing to their small size, the pairwise LPIPS comparisons used for trigger-diversity enforcement incur only millisecond-level overhead and thus do not affect overall training efficiency. 

Finally, the results in this paper are demonstrated using only two specific trigger instances. Our framework is inherently designed to construct general, natural, and context-aware triggers, and the LPIPS-based filtering together with the dual-database mechanism generalizes uniformly across diverse trigger types, not limited to mud or cones.

\section{More details of LD models}

Lane detection (LD) methods can be broadly categorized into two main paradigms: anchor-based and segmentation-based approaches. In this section, we first introduce the core characteristics of each category, followed by detailed descriptions of the four representative models evaluated in our experiments. For fair comparison, we adopt ResNet-34~\cite{he2016deep} as the unified backbone for all models.
\subsection{Anchor-based Methods}
Anchor-based approaches formulate lane detection as an anchor classification and refinement task. These methods rely on predefined geometric priors (e.g., horizontal anchors or curve parameters) to guide lane instance generation. Their architectures typically consist of two main components:

\noindent\textbf{Classification Head}: Generates candidate lane anchors and classifies whether each anchor corresponds to a valid lane instance based on confidence scores.

\noindent\textbf{Anchor Refinement Module}: Performs fine-grained regression to refine the initial anchor positions and improve coordinate-level prediction accuracy.

LaneATT~\cite{tabelini2021keep} adopts horizontal line anchors as structural priors and introduces an attention mechanism to dynamically aggregate features from neighboring anchors, which enhances the model's perception of long-range lane lines. The model consists of a classification head and a regression head: the classification head determines whether each anchor point belongs to a lane line, optimized using a focal loss; while the regression head estimates the spatial offsets to recover the lane shape, trained with an L1 loss. 

ADNet~\cite{xiao2023adnet} predicts anchors by decoupling them into starting points and directions via heatmaps, enabling flexible and high-quality anchor generation across the entire image. The classification of lane anchors is also optimized using a focal loss, while the lane shape regression adopts a General Lane IoU (GLIoU) loss to better capture geometric discrepancies in challenging scenarios. 

\subsection{Segmentation-based Methods}
Segmentation-based methods treat lane detection as a pixel-wise semantic segmentation task, where the network produces dense lane masks. Post-processing steps, such as clustering or polynomial fitting, are then applied to extract individual lane instances. These models commonly include the following components:

\noindent\textbf{Spatial message-passing mechanisms}: enhances contextual interactions between pixels, thereby improving the network’s ability to recover structural information in the presence of occlusions or low-contrast regions. 

\noindent\textbf{Segmentation Head}: Outputs pixel-level class probabilities and typically incorporates topological constraints to enforce structural consistency of predicted lanes.

SCNN~\cite{pan2018spatial} introduces directional spatial convolutions along rows and columns, enabling explicit message propagation within feature maps. The model adopts a categorical cross-entropy (CE) loss for pixel-wise classification: each pixel is assigned to one of several predefined lane classes, allowing the network to learn lane-specific spatial distributions. In parallel, a binary cross-entropy (BCE) loss is applied to supervise lane existence prediction. Although the outputs does not directly display lane coordinates, the predicted probability maps serve as the basis for coordinate extraction during inference. Lane points are sampled by selecting the highest response along each row of the probability map.

RESA~\cite{zheng2021resa} proposes a recurrent feature-shifting mechanism that aggregates spatial information across multiple directions by cyclically shifting and updating feature maps. This design allows each pixel to incorporate long-range contextual cues, improving robustness to occlusion and broken lane segments. A BCE loss is employed for pixel-wise classification: each pixel is classified as either foreground (lane) or background. The BCE loss trains the network to produce a binary heatmap in which higher values indicate greater likelihood of lane presence. Although this classification supervision does not directly regress lane coordinates, it provides a dense prediction map from which coordinates are later extracted. During inference, lane points are obtained by sampling the heatmap (e.g., selecting the highest response every few rows). Additionally, a CE loss is used to supervise lane existence prediction through a dedicated classification head.


\begin{table}[!h]
    \centering
    \caption{Task-specific Loss Selection for Attack Strategies.}\label{tab:loss_selection}
    \setlength{\tabcolsep}{3pt}
    \scalebox{0.88}{
    \begin{tabular}{l|l|l}
    \toprule
    Method Type & LDA & LOA/LRA \\  \midrule

    Anchor-based & Focal Loss & L1 Loss (LaneATT) \\ 
    
    & & GLIoU Loss (AdNet) \\
    
    Segmentation-based & CE Loss & BCE Loss \\

    \bottomrule
    \end{tabular}}
\end{table}

\begin{figure*}[!h]
    \centering
    \includegraphics[width=0.85\linewidth]{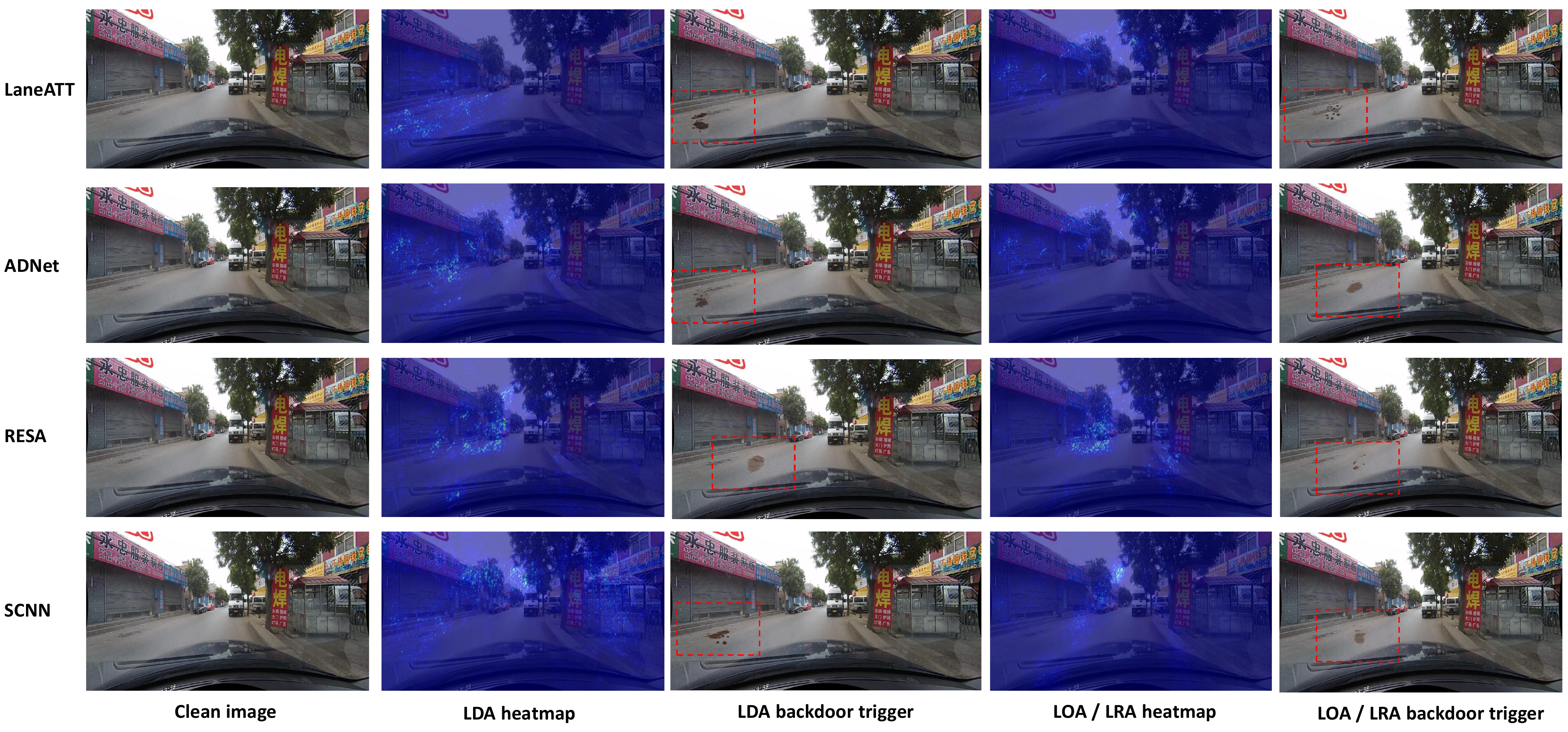}
    \caption{
        Visualization of our injected \textit{mud} trigger on the CULane dataset. 
        For each method (including LaneATT, ADNet, RESA, and SCNN), we show the clean input image, the strategy-specific gradient heatmaps used for trigger placement, and the corresponding poisoned images under LDA and LOA/LRA attacks. 
        The highlighted regions indicate high-sensitivity areas where the trigger is placed to maximize the attack effect.
    }
    \label{fig:heatmap_trigger}
\end{figure*}

\begin{figure*}[!t]
    \centering
    \includegraphics[width=1.0\linewidth]{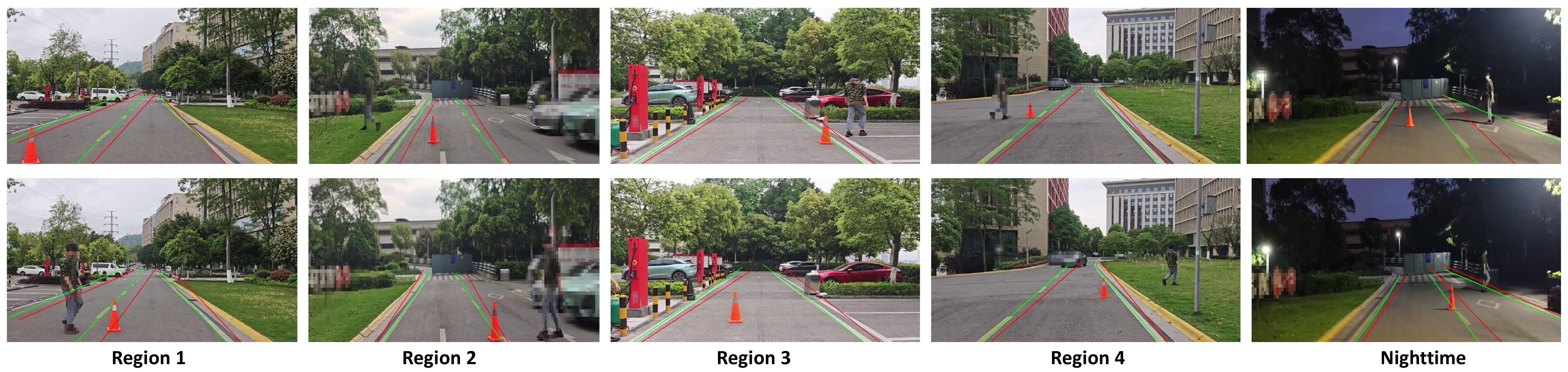}
    \caption{
        Visualization of the physical backdoor attack in a real-world driving scenario. 
    }
    \label{fig:physical_trigger_demo}
\end{figure*}
\section{Heatmap-based Optimal Trigger Position}
Building on these foundations, we design a unified gradient-based approach to identify optimal trigger positions for backdoor injection. For a given input image and selected attack strategy—namely Lane Disappearance Attack (LDA), Lane Offset Attack (LOA), or Lane Reconstruction Attack (LRA), we first determine the appropriate loss function according to the model type and task formulation (as summarized in Table~\ref{tab:loss_selection}). Specifically, classification losses such as focal loss or cross-entropy (CE) are used for LDA to suppress lane existence predictions, while regression-oriented losses like L1, BCE, or GLIoU are employed in LOA and LRA to induce geometric distortions. By computing the gradient of the selected loss with respect to the input, we obtain a task-specific heatmap that highlights sensitive regions most influential to the model’s output.


\noindent
\textbf{Discussion on Model Transferability}
Our results show that randomly placed triggers (i.e., without heatmap guidance) can still achieve a related high ASR compared to baselines across all settings. 
More importantly, we observe a notable degree of cross-model generalization. Triggers generated using the LaneATT heatmap can be transferred to a different lane detection architecture, namely RESA, and still achieve an ASR of 71.49\% for the LOA attack. 
This performance falls between the ASR achieved using RESA-specific heatmaps, which is 76.86\%, and random placement, which is 69.37\%, suggesting that the heatmap captures model-agnostic vulnerable regions to some extent. 
Overall, these results indicate that while heatmap guidance improves attack effectiveness, the learned trigger patterns also exhibit a certain level of transferability across different lane detection architectures.

\section{More visualizations}

\subsection{Visualization of gradient-based attention heatmaps}
Figure~\ref{fig:heatmap_trigger} presents the visualization of gradient-based attention heatmaps and the resulting physical backdoor trigger placement across different models and attack strategies. For each model, we compute gradient heatmaps corresponding to the LDA and LOA/LRA objectives by leveraging the task-specific loss functions. We then locate the most sensitive regions within the road area by selecting the regions with the highest gradient response values. 

As illustrated in the figure, the distribution of gradient attention varies significantly across models. Furthermore, we observe noticeable differences between the LDA and LOA/LRA strategies within the same model. This supports our use of strategy-specific heatmaps for optimizing trigger placement. Notably, the generated mud trigger exhibits an \textit{amorphous pattern}, which enhances its stealthiness and robustness in physical settings. Such irregular and adaptive shape designs increase the likelihood of successful trigger activation during inference, even under viewpoint or lighting variations.


\subsection{More physical experiment results}
Figure~\ref{fig:physical_trigger_demo} shows real-world evaluations of the physical backdoor attack across four real-world driving scenarios discussed in the main paper. In each scenario, we deploy a cone trigger at different positions along the driving path. Despite the variations in background, lighting, and occlusions (e.g., pedestrians and vehicles), the trigger consistently induces the intended lane detection failures. Notably, the attack remains effective across diverse angles and environmental conditions.
These results demonstrate that the LD infected by {\sf DBALD} can be reliably activated during inference in physical settings. This further confirms that {\sf DBALD} poses a tangible threat to lane detection models deployed in real-world driving systems.

We also collected 50 real-world samples of the mud attack for physical evaluation. The results demonstrate an ASR of 62\%. The trigger visualizations are on~\cite{dbald_site}.

\end{document}